\documentclass{bioinfoArXiv} 
\copyrightyear{2017}
\pubyear{2017}
\usepackage{verbatim} 
\usepackage{amssymb}
\usepackage{amsmath}
\usepackage{algorithm2e}
\usepackage{verbatim} 
\usepackage{epstopdf}
\usepackage{setspace}
\usepackage{amsthm}
\usepackage{rotating}
\usepackage[T1]{fontenc}
\usepackage{subfig}
\newtheorem{thm}{Theorem}[section]

\begin{document}
\firstpage{1}
\title[Encoding DNA sequences by integer CGR]{Encoding DNA sequences by integer chaos game representation}
\author[Yin, C.]{Changchuan Yin \footnote{To whom correspondence should be addressed. Email: cyin1@uic.edu}}
\address{Department of Mathematics, Statistics and Computer Science\\ The University of Illinois at Chicago, Chicago, IL 60607-7045, USA\\}

\history{} %Received on XXXXX; revised on XXXXX; accepted on XXXXX}
\editor{} %Associate Editor: XXXXXXX}

\maketitle

\begin{abstract}
\section{Motivation:}
DNA sequences are fundamental for encoding genetic information. The genetic information may be understood not only by symbolic sequences but also from the hidden signals inside the sequences. The symbolic sequences need to be transformed into numerical sequences so the hidden signals can be revealed by signal processing techniques. All current transformation methods encode DNA sequences into numerical values of the same length. These representations have limitations in the applications of genomic signal compression, encryption, and steganography.

\section{Results:}
We propose an integer chaos game representation (iCGR) of DNA sequences and a lossless encoding method DNA sequences by the iCGR. In the iCGR method, a DNA sequence is represented by the iterated function of the nucleotides and their positions in the sequence. Then the DNA sequence can be uniquely encoded and recovered using three integers from iCGR. One integer is the sequence length and the other two integers represent the accumulated distributions of nucleotides in the sequence. The integer encoding scheme can compress a DNA sequence by 2 bits per nucleotide. The integer representation of DNA sequences provides a prospective tool for sequence analysis and operations.

\section{Availability:}
The Python programs in this study are freely available to the public at https://github.com/cyinbox/iCGR

\section{Key words:}
DNA sequence, CGR, encoding, decoding, compression
\section{Contact:} \href{cyin1@uic.edu}{cyin1@uic.edu}

\end{abstract}

\section{Introduction}
\label{Introduction}
%1026
In recent years the Next Generation Sequencing (NGS) techniques have resulted in massive DNA and protein sequences. There are strong demands for efficiently analyzing these genomic sequences. A DNA sequence consists of four types of nucleotides: Adenine (A), Guanine (G), Thymine (T) and Cytosine (C). DNA sequence analysis requires conversion of a symbolic sequence to a numerical sequence so that intrinsic patterns and characters can be characterized by digital signal processing approaches \citep{anastassiou2000frequency,mendizabal2017dna,yin2008numerical,yin2016periodic}. Numerical representations of DNA sequences are also essential to genome comparison, compression, encryption, and steganography.   

%1029
An effective numerical representation must be able to capture all significant properties of the biological reality without introducing any spurious effects. Currently, the most commonly used encoding method is the Voss 4D binary indicator representations \citep{felsenstein1982efficient,voss1992evolution}, which has been used in protein-coding prediction, similarity analysis, and periodicity detection in genomes. However, the Voss 4D method and DNA sequence mapping are not one-to-one. In 1990, Jeffrey first proposed a numerical and graphical Chaos Game Representation (CGR) of a DNA sequence \citep{jeffrey1990chaos}. The CGR is generated in a square with the four vertices for the nucleotides A, C, G, and T, respectively. In the CGR graph, the first is placed halfway between the center of the square and the vertex corresponding to the first nucleotide of the DNA sequence and successive points are generated halfway between the previous point and the vertex representing the nucleotide being plotted. An important feature of the CGR is that the value of any point in CGR contains the historical information of the preceding sequence and visually displays all subsequent frequencies of a given DNA sequence. The CGR preserves all statistical properties of DNA sequences and allows investigation of both local and global patterns in DNA sequences, visually revealing previously hidden sequence structures. The CGR was then developed for k-mer counting and referred to frequency CGR, which renders a unique 2D image signature for a genome sequence. Because CGR has a remarkable ability to differentiate between genetic sequences belonging to different species, and it has thus been proposed as a genomic signature \citep{deschavanne1999genomic,almeida2001analysis}. Due to the character of information preservation of CGR, it has been applied in different research domains including similarity analysis of genomes \citep{stan2010similarity,kari2015mapping,joseph2006chaos,hoang2016numerical}, detection of hidden periodicity signal in genomes \citep{messaoudi2014building}. However, all existing numerically representation methods of DNA sequences produce a list of values of the same length of DNA sequences, and these types of representations cannot be directly used for storing, compressing, encrypting, and aligning DNA sequences. 

%1122
In this paper, we propose an integer chaos game representation (iCGR) of DNA sequences, in which nucleotides of DNA sequences are represented by iterated integer functions. Using iCGR, a DNA sequence can be uniquely encoded and recovered by three integers. One of the integers is the length of the DNA sequence, and the other two integers are determined by the type and positions of nucleotides in the DNA sequence. One application of the encoding is to compress DNA sequences. The result shows that 2 bits are required for storing a nucleotide symbol in integer encoding, whereas the common character representation of a nucleotide needs 8 bits. The proposed method will have wide applications in NGS sequence analysis.

\section{Methods and Algorithms}
\subsection{Chaos Game Representation (CGR) of DNA sequences}
%1027
CGR is an iterative mapping and scale-independent representation for geometric representation of DNA sequences \citep{jeffrey1990chaos}. The CGR space can be viewed as a continuous reference system, where all possible sequences of any length occupy a unique position. The position is produced by the four possible nucleotides, which are treated as vertices of a unit binary square since a DNA sequence can be treated formally as a string of the four letters A, C, G, and T. In this study, we redesign the CGR corners of four nucleotides so the relationship between two nucleotides can be reflected by the distances of the CGR corners. The CGR vertices are assigned to the four nucleotides as $A=(1, 1), T=(-1, 1), C=(-1, -1)$, and $G=(1,-1)$ (Fig.1(a)). The CGR coordinates are calculated iteratively by moving a pointer to half the distance between the previous position and the current binary representation (Algorithm 1). For example, if the next nucleotide on the DNA sequence is G, then a point is plotted halfway between the previous point and the $G$ corner. The resulting graphic is planar, thus we call it as classical CGR representation. A CGR example of a short DNA sequence, TAGCA, is illustrated in Fig.1(b). The CGR representation of Human mitochondrial genome is shown in Fig.2. The CGR of Human mitochondrial genome reveals the fractal patterns within the genome. 

%1109 CGR3D/CGR1D_Plot1.m,CGR1D_11062017.m
\begin{figure}[tbp]
      	\centering
        \subfloat[]{\includegraphics[width=3.0in]{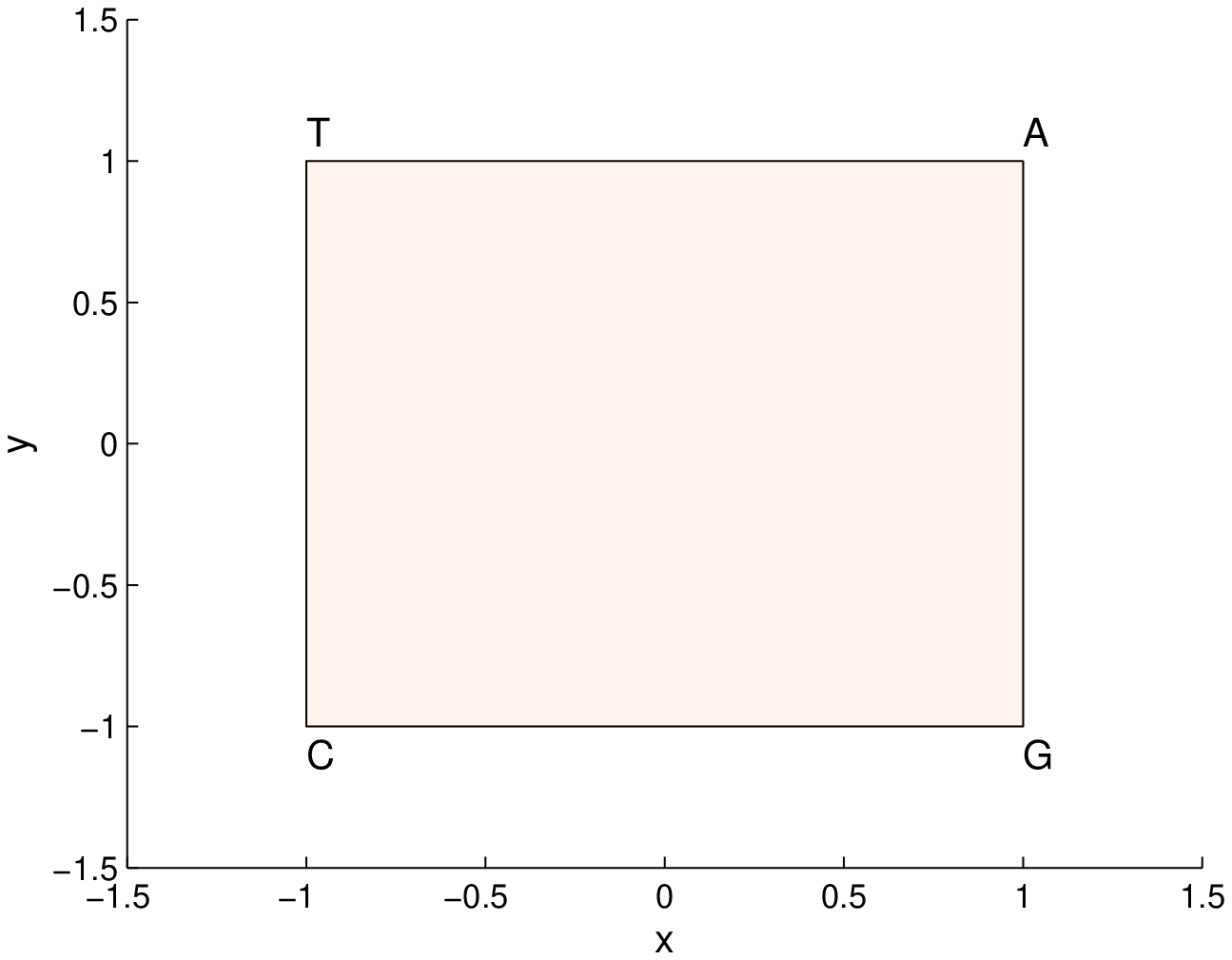}}\quad
      	\subfloat[]{\includegraphics[width=3.0in]{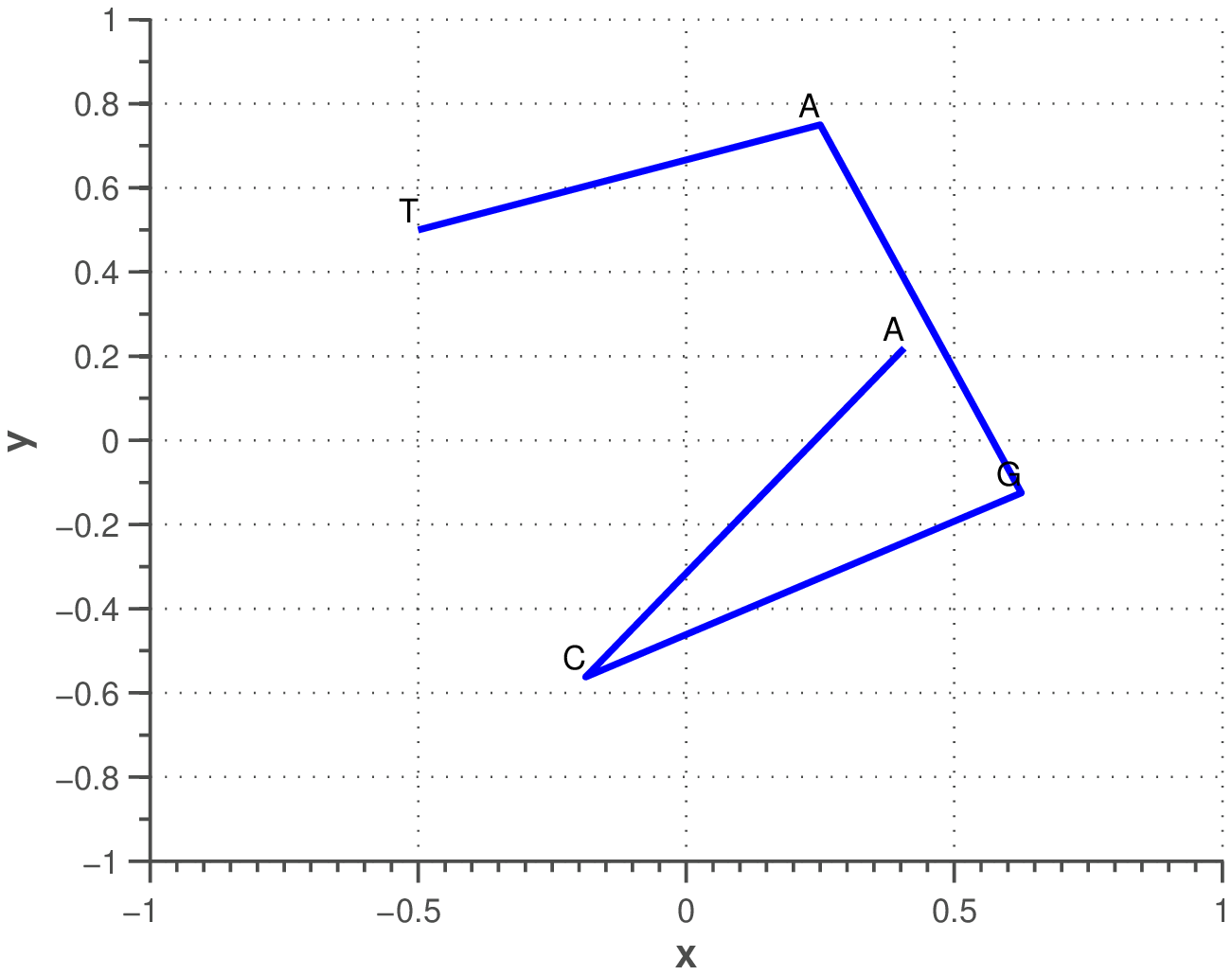}} %\quad
      	\caption{(a) Numerical representation of four nucleotides A, T, C, and G in CGR. (b) The GCR graph of the DNA sequence, TAGCA. The corresponding CGR coordinates are: $T_1(-0.5000,0.5000), A_2(0.2500,0.7500), G_3(0.6250,-0.1250),\\ C_4(-0.1875,-0.5625), A_5(0.4063,0.2188)$.}
      \end{figure}
%1110
 \begin{algorithm} %[H]
  \SetAlgoLined
  \KwIn{A DNA sequence $S$ of length $n$}
  \KwOut{List of 2D coordinates of the DNA sequence, $p_i (x ,y ),i=1,2,\cdots,n$}
  \textbf{Step:}
 \begin{enumerate}
   \item Create a square with four corners
    $A=(1, 1), T=(-1, 1), C=(-1, -1)$, and $G=(1, -1)$, representing the four nucleotides $A, T, C$, and $G$, respectively. These four points are denoted as CGR corners $\alpha  \in \{ A,T,C,G\}$.
    \item Initialize the first CGR coordinate based on the first nucleotide, $S(1)$, of the DNA sequence.\\
  \begin{equation} 
% MathType!MTEF!2!1!+-
% feaafiart1ev1aaatCvAUfeBSjuyZL2yd9gzLbvyNv2CaerbuLwBLn
% hiov2DGi1BTfMBaeXatLxBI9gBaerbd9wDYLwzYbItLDharqqtubsr
% 4rNCHbWexLMBbXgBd9gzLbvyNv2CaeHbl7mZLdGeaGqiVCI8FfYJH8
% YrFfeuY-Hhbbf9v8qqaqFr0xc9pk0xbba9q8WqFfeaY-biLkVcLq-J
% Hqpepeea0-as0Fb9pgeaYRXxe9vr0-vr0-vqpWqaaeaabiGaciaaca
% qabeaadaqaaqaafaGceaqabeaacaWGWbWaaSbaaSqaaiaaigdacaGG
% SaGaamiEaaqabaGccqGH9aqpdaWcaaqaaiaaigdaaeaacaaIYaaaai
% abeg7aHnaaBaaaleaacaaIXaGaaiilaiaadIhaaeqaaaGcbaGaamiC
% amaaBaaaleaacaaIXaGaaiilaiaadMhaaeqaaOGaeyypa0ZaaSaaae
% aacaaIXaaabaGaaGOmaaaacqaHXoqydaWgaaWcbaGaaGymaiaacYca
% caWG5baabeaaaOqaaiabeg7aHnaaBaaaleaacaaIXaaabeaakiabg2
% da9iaadofacaGGOaGaaGymaiaacMcacaGGSaGaeqySde2aaSbaaSqa
% aiaaigdaaeqaaOGaeyicI4Saai4EaiaadgeacaGGSaGaamivaiaacY
% cacaWGdbGaaiilaiaadEeacaGG9baaaaa!665A!
%\[
\begin{gathered}
  p_{1,x}  = \frac{1}
{2}\alpha _{1,x}  \hfill \\
  p_{1,y}  = \frac{1}
{2}\alpha _{1,y}  \hfill \\
  \alpha _1  = S(1),\alpha _1  \in \{ A,T,C,G\}  \hfill \\ 
\end{gathered} 
%\]
\end{equation}
   \item Compute the CGR coordinate of current nucleotide, $S(i)$, of the DNA sequence as the midpoint of previous coordinate and the CGR corner for this nucleotide.\\
     \begin{equation}
 % MathType!MTEF!2!1!+-
 % feaafiart1ev1aaatCvAUfeBSjuyZL2yd9gzLbvyNv2CaerbuLwBLn
 % hiov2DGi1BTfMBaeXatLxBI9gBaerbd9wDYLwzYbItLDharqqtubsr
 % 4rNCHbWexLMBbXgBd9gzLbvyNv2CaeHbl7mZLdGeaGqiVCI8FfYJH8
 % YrFfeuY-Hhbbf9v8qqaqFr0xc9pk0xbba9q8WqFfeaY-biLkVcLq-J
 % Hqpepeea0-as0Fb9pgeaYRXxe9vr0-vr0-vqpWqaaeaabiGaciaaca
 % qabeaadaqaaqaafaGceaqabeaacaWGWbWaaSbaaSqaaiaadMgacaGG
 % SaGaamiEaaqabaGccqGH9aqpdaWcaaqaaiaaigdaaeaacaaIYaaaai
 % aacIcacaWGWbWaaSbaaSqaaiaadMgacqGHsislcaaIXaGaaiilaiaa
 % dIhaaeqaaOGaey4kaSIaeqySde2aaSbaaSqaaiaadMgacaGGSaGaam
 % iEaaqabaGccaGGPaaabaGaamiCamaaBaaaleaacaWGPbGaaiilaiaa
 % dMhaaeqaaOGaeyypa0ZaaSaaaeaacaaIXaaabaGaaGOmaaaacaGGOa
 % GaamiCamaaBaaaleaacaWGPbGaeyOeI0IaaGymaiaacYcacaWG5baa
 % beaakiabgUcaRiabeg7aHnaaBaaaleaacaWGPbGaaiilaiaadMhaae
 % qaaOGaaiykaaqaaiabeg7aHnaaBaaaleaacaWGPbaabeaakiabg2da
 % 9iaadofacaGGOaGaamyAaiaacMcacaGGSaGaeqySde2aaSbaaSqaai
 % aadMgaaeqaaOGaeyicI4Saai4EaiaadgeacaGGSaGaamivaiaacYca
 % caWGdbGaaiilaiaadEeacaGG9baabaGaamyAaiabg2da9iaaikdaca
 % GGSaGaaG4maiaacYcacqWIMaYscaGGSaGaamOBaaaaaa!7EA5!
 %\[
 \begin{gathered}
   p_{i,x}  = \frac{1}
 {2}(p_{i - 1,x}  + \alpha _{i,x} ) \hfill \\
   p_{i,y}  = \frac{1}
 {2}(p_{i - 1,y}  + \alpha _{i,y} ) \hfill \\
   \alpha _i  = S(i),\alpha _i  \in \{ A,T,C,G\}  \hfill \\
   i = 2,3, \ldots ,n \hfill \\ 
 \end{gathered} 
 %\]
 \end{equation}
         
\end{enumerate}
  \caption{Algorithm for computing the CGR of a DNA sequence.}
\end{algorithm}
%CGR1D_surfScatter_11102017.m
\begin{figure}[tbp]
    \centering
    {\includegraphics[width=3.75in]{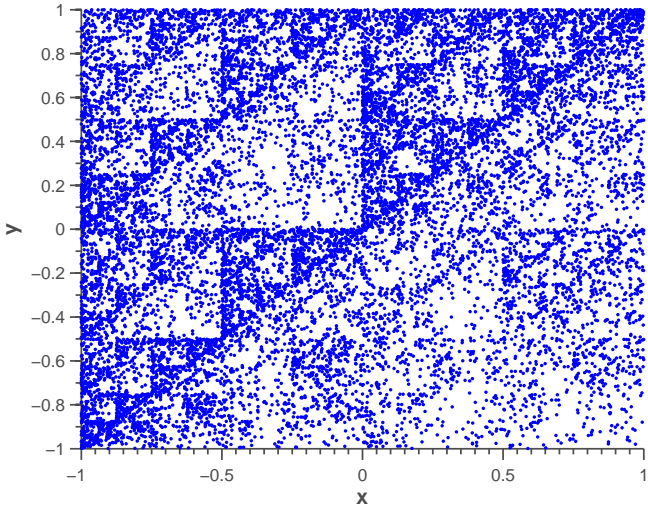}}
    \caption{The CGR graph of Human mitochondrial genome (GenBank access number: D38116).}
 \end{figure}
\subsection{Encoding DNA sequences by integer CGR} 
%1105
From the definition of CGR, we notice that the CGR coordinates of current nucleotide is determined by the CGR coordinates of the preceding nucleotide and the fixed CGR corner coordinates of the current nucleotide. According to this recursive relationship, we may get the final CGR coordinates of a DNA sequence. The CGR theorem suggests that the final coordinates contain the full DNA sequence information \citep{jeffrey1990chaos}. However, due to the floating-point errors in the computation of CGR, DNA sequences cannot be fully recovered by the final CGR coordinates.

%1108
To address the floating-point errors when encoding DNA sequences in the original CGR scheme, here we redefine a new CGR coordinate relationship as in Equations (3) and (4). Instead of taking the midpoint of the preceding position and the current CGR corner as in original CGR \citep{jeffrey1990chaos}, the current position the new CGR schema is the sum of the preceding coordinate and exponential of two of the CGR corner (Equations (4)). As an example, the integer CGR of a short DNA sequence, TAGCA, is illustrated in Fig.3. It should be noted that the proposed CGR mapping of DNA sequences is different from the original CGR. In the proposed CGR mapping, the coordinates of DNA sequences are integers and can extend all the 2-dimensional space, while regular CGR coordinates are float numbers and are limited to the unit square. Thus, we may consider that the proposed integer CGR is an open mapping, and the original CGR is a closed mapping. Although the original CGR coordinates are determined by the DNA sequences from theoretical analysis, due to floating-point errors, the original CGR cannot recover a DNA sequence when the length of the sequence is longer than 32 bp. Our proposed CGR mapping results in integer coordinates without the floating-point errors, therefore, the integer CGR can recover a full DNA sequence when the length of the sequence is less than 1024 bp. This is the significant advantage of our proposed integer CGR mapping.

%1109 CGR3D/iCGR1D_11062017.m
\begin{figure}[tbp]
      	\centering
        {\includegraphics[width=3.0in]{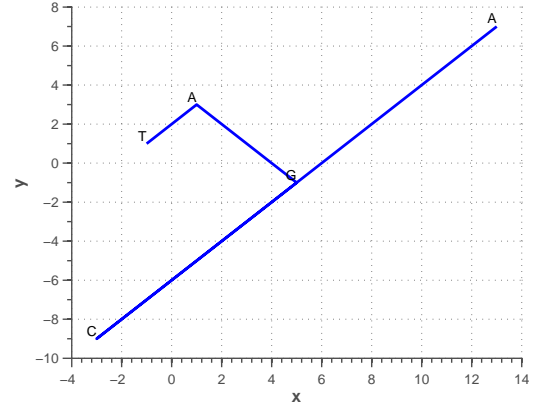}} 
      	\caption{Integer CGR encoding the DNA sequence, TAGCA. The corresponding iCGR coordinates are: $T_1(-1,1), A_2(1,3), G_3(5,-1), C_4(-3,-9), A_5(13,7)$ }.
      \end{figure} 

We propose here an integer CGR encoding algorithm for a DNA sequence. Using this algorithm, a DNA sequence can be uniquely represented by three numbers: The length of the sequence and the two integers of the final CGR coordinate of the DNA sequence. These integers contain all the DNA sequence information and can recover the sequence reversely. Encoding a DNA sequence into three integers by integer CGR is as follows (Equation(3) and algorithm 2). We first initialize the CGR coordinate at the first position of a DNA sequence using the CGR corner coordinate, and then the following CGR coordinate is computed based on the preceding coordinate and the nucleotide at this position.
\begin{equation}
% MathType!MTEF!2!1!+-
% feaafiart1ev1aaatCvAUfeBSjuyZL2yd9gzLbvyNv2CaerbuLwBLn
% hiov2DGi1BTfMBaeXatLxBI9gBaerbd9wDYLwzYbItLDharqqtubsr
% 4rNCHbWexLMBbXgBd9gzLbvyNv2CaeHbl7mZLdGeaGqiVCI8FfYJH8
% YrFfeuY-Hhbbf9v8qqaqFr0xc9pk0xbba9q8WqFfeaY-biLkVcLq-J
% Hqpepeea0-as0Fb9pgeaYRXxe9vr0-vr0-vqpWqaaeaabiGaciaaca
% qabeaadaqaaqaafaGceaqabeaacaWGWbWaaSbaaSqaaiaaigdacaGG
% SaGaamiEaaqabaGccqGH9aqpcqaHXoqydaWgaaWcbaGaaGymaiaacY
% cacaWG4baabeaaaOqaaiaadchadaWgaaWcbaGaaGymaiaacYcacaWG
% 5baabeaakiabg2da9iabeg7aHnaaBaaaleaacaaIXaGaaiilaiaadM
% haaeqaaaGcbaGaeqySde2aaSbaaSqaaiaaigdaaeqaaOGaeyypa0Ja
% am4uaiaacIcacaaIXaGaaiykaiaacYcacqaHXoqydaWgaaWcbaGaaG
% ymaaqabaGccqGHiiIZcaGG7bGaamyqaiaacYcacaWGubGaaiilaiaa
% doeacaGGSaGaam4raiaac2haaaaa!634C!
%\[
\begin{gathered}
  p_{1,x}  = \alpha _{1,x}  \hfill \\
  p_{1,y}  = \alpha _{1,y}  \hfill \\
  \alpha _1  = S(1),\alpha _1  \in \{ A,T,C,G\}  \hfill \\ 
\end{gathered} 
%\]
\end{equation}

%1122
Then we can get all the current integer CGR coordinate at position $i$ based on the preceding coordinate at position $i-1$ and the nucleotide at position $i$.
\begin{equation}
% MathType!MTEF!2!1!+-
% feaafiart1ev1aaatCvAUfeBSjuyZL2yd9gzLbvyNv2CaerbuLwBLn
% hiov2DGi1BTfMBaeXatLxBI9gBaerbd9wDYLwzYbItLDharqqtubsr
% 4rNCHbWexLMBbXgBd9gzLbvyNv2CaeHbl7mZLdGeaGqiVCI8FfYJH8
% YrFfeuY-Hhbbf9v8qqaqFr0xc9pk0xbba9q8WqFfeaY-biLkVcLq-J
% Hqpepeea0-as0Fb9pgeaYRXxe9vr0-vr0-vqpWqaaeaabiGaciaaca
% qabeaadaqaaqaafaGceaqabeaacaWGWbWaaSbaaSqaaiaadMgacaGG
% SaGaamiEaaqabaGccqGH9aqpcaWGWbWaaSbaaSqaaiaadMgacqGHsi
% slcaaIXaGaaiilaiaadIhaaeqaaOGaey4kaSIaaGOmamaaCaaaleqa
% baGaamyAaiabgkHiTiaaigdaaaGccqaHXoqydaWgaaWcbaGaamyAai
% aacYcacaWG4baabeaaaOqaaiaadchadaWgaaWcbaGaamyAaiaacYca
% caWG5baabeaakiabg2da9iaadchadaWgaaWcbaGaamyAaiabgkHiTi
% aaigdacaGGSaGaamyEaaqabaGccqGHRaWkcaaIYaWaaWbaaSqabeaa
% caWGPbGaeyOeI0IaaGymaaaakiabeg7aHnaaBaaaleaacaWGPbGaai
% ilaiaadMhaaeqaaaGcbaGaamyyamaaBaaaleaacaWGPbaabeaakiab
% g2da9iaadofacaGGOaGaamyAaiaacMcacaGGSaGaamyyamaaBaaale
% aacaWGPbaabeaakiabgIGiolaacUhacaWGbbGaaiilaiaadsfacaGG
% SaGaam4qaiaacYcacaWGhbGaaiyFaaqaaiaadMgacqGH9aqpcaaIYa
% GaaiilaiablAciljaacYcacaWGUbaaaaa!7D18!
%\[
\begin{gathered}
  p_{i,x}  = p_{i - 1,x}  + 2^{i - 1} \alpha _{i,x}  \hfill \\
  p_{i,y}  = p_{i - 1,y}  + 2^{i - 1} \alpha _{i,y}  \hfill \\
  a_i  = S(i),a_i  \in \{ A,T,C,G\}  \hfill \\
  i = 2, \ldots ,n \hfill \\ 
\end{gathered} 
%\]
\end{equation}

%1118
From the recursive relation (Equation(4)), we may prove that the integer CGR coordinates are the sum of the product of position exponents and nucleotide types (Equation(5)). When a DNA sequence of length $n$ is finally encoded by the three integer numbers $(n,p_{n,x},p_{n,x})$ (tri-integers) through the integer CGR, these three integer reflect the accumulative distribution of nucleotides along the sequence. Because the final encoding tri-integers hold all the sequence information, we propose to use the encoding tri-integers as the signature of a DNA sequence.
\begin{equation}
% MathType!MTEF!2!1!+-
% feaafiart1ev1aaatCvAUfeBSjuyZL2yd9gzLbvyNv2CaerbuLwBLn
% hiov2DGi1BTfMBaeXatLxBI9gBaerbd9wDYLwzYbItLDharqqtubsr
% 4rNCHbWexLMBbXgBd9gzLbvyNv2CaeHbl7mZLdGeaGqiVCI8FfYJH8
% YrFfeuY-Hhbbf9v8qqaqFr0xc9pk0xbba9q8WqFfeaY-biLkVcLq-J
% Hqpepeea0-as0Fb9pgeaYRXxe9vr0-vr0-vqpWqaaeaabiGaciaaca
% qabeaadaqaaqaafaGceaqabeaacaWGWbWaaSbaaSqaaiaad6gacaGG
% SaGaamiEaaqabaGccqGH9aqpdaaeWbqaaiaaikdadaahaaWcbeqaai
% aadMgacqGHsislcaaIXaaaaOGaeqySde2aaSbaaSqaaiaadMgacaGG
% SaGaamiEaaqabaaabaGaamyAaiabg2da9iaaigdaaeaacaWGUbaani
% abggHiLdaakeaacaWGWbWaaSbaaSqaaiaad6gacaGGSaGaamyEaaqa
% baGccqGH9aqpdaaeWbqaaiaaikdadaahaaWcbeqaaiaadMgacqGHsi
% slcaaIXaaaaOGaeqySde2aaSbaaSqaaiaadMgacaGGSaGaamyEaaqa
% baaabaGaamyAaiabg2da9iaaigdaaeaacaWGUbaaniabggHiLdaake
% aacqaHXoqydaWgaaWcbaGaamyAaaqabaGccqGH9aqpcaWGtbGaaiik
% aiaadMgacaGGPaaabaGaamyAaiabg2da9iaaigdacaGGSaGaaGOmai
% aacYcacqWIVlctcaGGSaGaamOBaaaaaa!73A2!
%\[
\begin{gathered}
  p_{n,x}  = \sum\limits_{i = 1}^n {2^{i - 1} \alpha _{i,x} }  \hfill \\
  p_{n,y}  = \sum\limits_{i = 1}^n {2^{i - 1} \alpha _{i,y} }  \hfill \\
  \alpha _i  = S(i) \hfill \\
  i = 1,2, \cdots ,n \hfill \\ 
\end{gathered} 
%\]
\end{equation}
\subsection{Integer decoding DNA sequences} 
%1117
We can prove that the sign of $p_{i}$ is the same as that of nucleotide $\alpha _{i}$. Therefore, the integer CGR coordinates can be used to determine the corresponding nucleotide types. The nucleotide at the position $i$ can be determined based on the integer CGR coordinate at this position $p(x_i,y_i)$ using the properties as in Equations (6) and (7). This can lead the full recovery of original sequence from the final tri-integers $(n, X, Y)$.
\begin{equation}
% MathType!MTEF!2!1!+-
% feaafiart1ev1aaatCvAUfeBSjuyZL2yd9gzLbvyNv2CaerbuLwBLn
% hiov2DGi1BTfMBaeXatLxBI9gBaerbd9wDYLwzYbItLDharqqtubsr
% 4rNCHbWexLMBbXgBd9gzLbvyNv2CaeHbl7mZLdGeaGqiVCI8FfYJH8
% YrFfeuY-Hhbbf9v8qqaqFr0xc9pk0xbba9q8WqFfeaY-biLkVcLq-J
% Hqpepeea0-as0Fb9pgeaYRXxe9vr0-vr0-vqpWqaaeaabiGaciaaca
% qabeaadaqaaqaafaGcbaGaeqySde2aaSbaaSqaaiaadMgaaeqaaOGa
% eyypa0ZaaiqaaqaabeqaaiaadgeacaGGSaGaaGjcVlaayIW7caaMi8
% Uaey4aIqIaaGjcVlaayIW7caGGOaGaamiCamaaBaaaleaacaWGPbGa
% aiilaiaadIhaaeqaaOGaeyOpa4JaaGimaiaayIW7caaMi8UaaGjcVl
% aacAcacaqGGaGaamiCamaaBaaaleaacaWGPbGaaiilaiaadMhaaeqa
% aOGaeyOpa4JaaGimaiaacMcaaeaacaqGubGaaeilaiaayIW7caaMi8
% UaaGjcVlaayIW7caaMi8Uaey4aIqIaaGjcVlaayIW7caGGOaGaamiC
% amaaBaaaleaacaWGPbGaaiilaiaadIhaaeqaaOGaeyipaWJaaGimai
% aayIW7caaMi8UaaGjcVlaacAcacaqGGaGaamiCamaaBaaaleaacaWG
% PbGaaiilaiaadMhaaeqaaOGaeyOpa4JaaGimaiaacMcaaeaacaWGdb
% GaaiilaiaayIW7caaMi8UaaGjcVlabgoGiKiaayIW7caaMi8Uaaiik
% aiaadchadaWgaaWcbaGaamyAaiaacYcacaWG4baabeaakiabgYda8i
% aaicdacaaMi8UaaGjcVlaayIW7caGGMaGaaeiiaiaadchadaWgaaWc
% baGaamyAaiaacYcacaWG5baabeaakiabgYda8iaaicdacaGGPaaaba
% Gaam4raiaacYcacaaMi8UaaGjcVlaayIW7cqGHdicjcaaMi8UaaGjc
% VlaacIcacaWGWbWaaSbaaSqaaiaadMgacaGGSaGaamiEaaqabaGccq
% GH+aGpcaaIWaGaaGjcVlaayIW7caaMi8UaaGjcVlaacAcacaqGGaGa
% amiCamaaBaaaleaacaWGPbGaaiilaiaadMhaaeqaaOGaeyipaWJaaG
% imaiaacMcaaaGaay5Eaaaaaa!BADC!
%\[
\alpha _i  = \left\{ \begin{gathered}
  A,{\kern 1pt} {\kern 1pt} {\kern 1pt} \exists {\kern 1pt} {\kern 1pt} (p_{i,x}  > 0{\kern 1pt} {\kern 1pt} {\kern 1pt} \& {\text{ }}p_{i,y}  > 0) \hfill \\
  {\text{T,}}{\kern 1pt} {\kern 1pt} {\kern 1pt} {\kern 1pt} {\kern 1pt} \exists {\kern 1pt} {\kern 1pt} (p_{i,x}  < 0{\kern 1pt} {\kern 1pt} {\kern 1pt} \& {\text{ }}p_{i,y}  > 0) \hfill \\
  C,{\kern 1pt} {\kern 1pt} {\kern 1pt} \exists {\kern 1pt} {\kern 1pt} (p_{i,x}  < 0{\kern 1pt} {\kern 1pt} {\kern 1pt} \& {\text{ }}p_{i,y}  < 0) \hfill \\
  G,{\kern 1pt} {\kern 1pt} {\kern 1pt} \exists {\kern 1pt} {\kern 1pt} (p_{i,x}  > 0{\kern 1pt} {\kern 1pt} {\kern 1pt} {\kern 1pt} \& {\text{ }}p_{i,y}  < 0) \hfill \\ 
\end{gathered}  \right.
%\]
\end{equation}

%1116
We have the following theorem on integer encoding DNA sequences.

\begin{thm}
When a DNA sequence of length $n$ is encoded by the CGR coordinates $(X, Y)$, the sequence information can be fully recovered from tri-integers $(n, X, Y)$. 
\end{thm}
%1102
After a DNA sequence of length $n$ is encoded recursively and represented by the encoding integers $(n,X,Y)$ of the last step of encoding. If a DNA sequence of length $n$ is encoded as tri-integers $(n, X, Y)$, where $X$ is the $x$ coordinate at position $n$, and $Y$ is the $y$ coordinate at position n. Thus we can decode and recover the original DNA sequence from these tri-integers $(n, X, Y)$. The first step in the decoding process is to determine the nucleotide $\alpha_n$ at the last position $n$. From Equation (6), we may determine the nucleotide $\alpha_n$. For example, if $X = -19, Y = -25$, then the nucleotide at $n$ is $A$. Then we can recover the second last coordinate $(p_{n - 1,x}, p_{n - 1,y} )$ from the last coordinate $(x_n , y_n )$ and the vertex coordinate of last nucleotide $\alpha_n$ (Equation (7)). After we get $(p_{n - 1,x}, p_{n - 1,y} )$, we may determine the nucleotide $\alpha_{n-1}$ at the position $n-1$ from $(p_{n - 1,x}, p_{n - 1,y})$ by Equation (6). Using this method, all the nucleotides at positions $i = n - 1, n-2, \ldots , 1$ can be recursively determined.  

%1126
By this encoding method, we can see that any DNA sequences that end with nucleotide A are encoded by two large integers in Quadrant I, those end with nucleotide T are in Quadrant II, those end with nucleotide C are in Quadrant III, and those end with nucleotide G are in Quadrant IV. From the locations of the sequences, we can determine the type of the last nucleotide of the sequences. 

%1115
After a DNA sequence is encoded into tri-integers by the integer CGR scheme, the sequence can be fully recovered from the tri-integers. To recover the DNA sequence from encoded tri-integers $(n,X,Y)$, we can first determine the last nucleotide according to Equation (6). Since the GCR corner coordinate $\alpha_n$ is known, we can use the following induction to obtain the CGR coordinate of $n-1$ position, and then determine the nucleotide by Equation (7). Using the iteration process, all the nucleotides at all the positions, $n-1, n-2, .... 1$, can be determined (Algorithm 3). 

%1125
The tri-integers can also detect single nucleotide mutation in a DNA sequence. For example, the mutation from A to T in a DNA sequence, of which the wild-type sequence is represented by tri-integer ($n, X, Y$), the mutation generates new tri-integers ($n, X_m, Y_m$). The difference of ($n, X, Y$) and ($n, X_m, Y_m$) can determine the single nucleotide mutation.

%1104 
It is noted that the encoding scheme can detect an error if the given tri-integers are not for a DNA sequence. In each step of decoding, the value of each nucleotide can be recovered. If the values are not 1/-1 pairs, then the given tri-integers are not for a DNA sequence. Therefore, when DNA sequences are encoded by the proposed integer CGR, if there is an error during data storage and transfer, the iCGR encoding and decoding method can detect this error at the location.

\begin{equation}
% MathType!MTEF!2!1!+-
% feaafiart1ev1aaatCvAUfeBSjuyZL2yd9gzLbvyNv2CaerbuLwBLn
% hiov2DGi1BTfMBaeXatLxBI9gBaerbd9wDYLwzYbItLDharqqtubsr
% 4rNCHbWexLMBbXgBd9gzLbvyNv2CaeHbl7mZLdGeaGqiVCI8FfYJH8
% YrFfeuY-Hhbbf9v8qqaqFr0xc9pk0xbba9q8WqFfeaY-biLkVcLq-J
% Hqpepeea0-as0Fb9pgeaYRXxe9vr0-vr0-vqpWqaaeaabiGaciaaca
% qabeaadaqaaqaafaGceaqabeaacaWGWbWaaSbaaSqaaiaadMgacqGH
% sislcaaIXaGaaiilaiaadIhaaeqaaOGaeyypa0JaamiCamaaBaaale
% aacaWGPbGaaiilaiaadIhaaeqaaOGaeyOeI0IaaGOmamaaCaaaleqa
% baGaamyAaaaakiabeg7aHnaaBaaaleaacaWGPbGaaiilaiaadIhaae
% qaaaGcbaGaamiCamaaBaaaleaacaWGPbGaeyOeI0IaaGymaiaacYca
% caWG5baabeaakiabg2da9iaadchadaWgaaWcbaGaamyAaiaacYcaca
% WG5baabeaakiabgkHiTiaaikdadaahaaWcbeqaaiaadMgaaaGccqaH
% XoqydaWgaaWcbaGaamyAaiaacYcacaWG5baabeaaaOqaaiaadMgacq
% GH9aqpcaaIYaGaaiilaiablAciljaacYcacaWGUbaaaaa!682D!
%\[
\begin{gathered}
  p_{i - 1,x}  = p_{i,x}  - 2^i \alpha _{i,x}  \hfill \\
  p_{i - 1,y}  = p_{i,y}  - 2^i \alpha _{i,y}  \hfill \\
  i = 2, \ldots ,n \hfill \\ 
\end{gathered} 
%\]
\end{equation}

\begin{algorithm} %[H]
 \SetAlgoLined
 \KwIn{A DNA sequence $S$ of length $n$}
 \KwOut{Tri-integers $(n,X,Y)$ representing the DNA sequence}
 \textbf{Step:}
 \begin{enumerate}
  \item Get the nucleotide coordinates $\alpha_{1,x},\alpha_{1,y}$ at position 1 from  based on Equation (3).
  \item Compute the x-coordinate at position $i$ from that at position $i-1$: $p_{i,x}  = p_{i - 1,x}  + 2^{i - 1} \alpha _{i,x}$.
  \item Compute the x-coordinate at position $i$ from that at position $i-1$: $p_{i,y}  = p_{i - 1,y}  + 2^{i - 1} \alpha _{i,y}$.
  \item Repeat steps 2 and 3 until $i=n$.
  \item When $i=n$, return $X=p_{n,x}$, $Y=p_{n,y}$.
 
 \end{enumerate}
 \caption{Encoding a DNA sequence by tri-integers $(n,X,Y)$.}
\end{algorithm}

%PythonScript/CGREncoder_111142017.py
\begin{algorithm} %[H]
 \SetAlgoLined
 \KwIn{Tri-integers $(n,X,Y)$}
 \KwOut{The DNA sequence that is encoded by tri-integers $(n,X,Y)$}
 \textbf{Step:}
 \begin{enumerate}
  \item Get the nucleotide $\alpha_n$ at position n from $(n,X,Y)$ based on Equation (3). 
  \item Compute the x-coordinate at position $i-1$ from that at position $i$: $p_{i - 1,x}  = p_{i,x}  - 2^{i - 1} \alpha _{i,x}$.
  \item Compute the y-coordinate at position $i-1$ from that at position $i$: $p_{i - 1,y}  = p_{i,y}  - 2^{i - 1} \alpha _{i,y}$.
  \item Get the nucleotide $\alpha_{i-1}$ at position $i-1$ from $p_{i - 1,x},p_{i - 1,y}$ based on equation (3).
  \item Repeat steps 2, 3 and 4 until $i=1$.
  \item Return the decoded nucleotide sequence of length n.
 \end{enumerate}
 \caption{Decoding a DNA sequence from tri-integers $(n,X,Y)$.}
\end{algorithm}

\section{Results}
%1120
In the encoding and decoding algorithms, we consider two classes of four different bases in DNA, the pyrimidines (Cytosine (C) and Thymine (T)), and the purines (Adenine (A) and Guanine (G)), and design new CGR corners. The algorithms use integer iteration so the relationship of nucleotide positions and encoding integers are one-to-one. We perform case studies for evaluating the effectiveness of the integer encoding and decoding algorithms in sequence representation and compression.

%\subsection{Integer encoding and decoding DNA sequences} 
%1102 PythonScript/CGREncoder_111142017.py 
An example of encoding a short DNA sequence of length 10 bp is illustrated in Table 1. Table 1 shows the encoded $p_{i,x}$ and $p_{i,y}$ at each position $i$. The two large integers in the final step for $n=10$ are -203 and 441. The DNA sequence can be encoded and stored by these integers, $(n = 10,X = -203,Y = 441)$. 
 \begin{table}[ht]
 	\caption{Encoding a DNA sequence of length 10 by tri-integers $(n,X,Y)$}
 	\centering % used for centering table
 	%\begin{center}
 	\begin{tabular}{l*{12}{c}r}
 		\hline\hline
 		DNA   & C & G & T & A & A & C & T & A & G & T \\
 		$i$   & 1 & 2 & 3 & 4 & 5 & 6 & 7 & 8 & 9 & 10 \\
 		$\alpha_{i,x}$ & -1 & 1 & -1  & 1 & 1 & -1 & -1 & 1 & 1 & -1 \\
 	    $\alpha_{i,y}$ & -1 & -1 & 1 & 1 &  1 & -1 & 1 & 1 & -1 & 1 \\
 		\hline
 		$ p_{i,x}$  & -1 & 1 & -3 & 5 & 21 & -11 & -75 & 53 & 309 & -203\\
 	    $ p_{i,y}$  & -1 & -3 & 1 & 9 & 25 & -7 & 57 & 185 & -71 & 441\\
 		\hline\hline
 	\end{tabular}
 	%\end{center}
 	\label{table:nonlin} 
 \end{table}
 
%1102 PythonScript/CGREncoder_111142017.py
An example of decoding a DNA sequence is illustrated in Table 2. Table 2 shows the encoded integers $p_{i,x}$ and $p_{i,y}$ and decoded nucleotide at each position $i$. The tri-integers $(n = 10, X = 659, Y = 783)$ are used for recovering the full sequence. The DNA sequence recovered by the tri-integers is 'ATTGCCGTAA'
 \begin{table}[ht]
 	\caption{Decoding a DNA sequence from tri-integers $(n=10, X=659, Y=783)$}
 	\centering % used for centering table
 	%\begin{center}
 	\begin{tabular}{l*{12}{c}r}
 		\hline\hline
 	    $i$   & 10 & 9 & 8 & 7 & 6 & 5 & 4 & 3 & 2 & 1 \\
 		\hline
 		$ p_{i,x}$  & 659 &147 &-109 & 19 & -45 & -13 & 3 & -5 & -1 & 1\\
 	    $ p_{i,y}$  & 783 &271 & 15 & -113 & -49 & -17 & -1 & 7 & 3 & 1\\
 	    Nucleotide  & A & A & T & G & C & C & G & T & T & A \\
 	     
 		\hline\hline
 	\end{tabular}
 	%\end{center}
 	\label{table:nonlin} 
 \end{table}
 
Another example of encoding is for encoding Homo sapiens globin gene (GenBank access number: HF583935). For each position of the sequence, we generate the encoding iCGR coordinates as shown in Fig.4. The final encoding tri-integers for this gene are $(171,20503394090813028165419285686907647331949312955\\96027, 
111945316267328672851214367944002385890505723\\6646823)$. 

%iCGR_globin_11162017,iCGR_microSatellite_11162017
  \begin{figure}[tbp]
      \centering
      %\subfloat[]{\includegraphics[width=3.5in]{iCGR_globin_11162017_fig1.eps}}
      \includegraphics[width=3.5in]{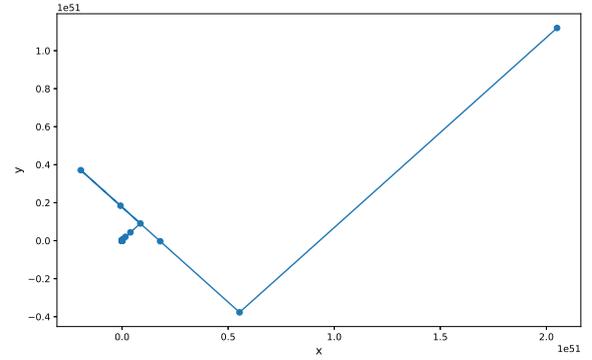}
      \caption{Integer encoding Homo sapiens globin gene (GenBank access number: HF583935).}
   \end{figure}

%1103
Using the encoding method, the average bits of the tri-integers per a nucleotide is 2.0, whereas symbolic DNA sequences need 8 bits per nucleotide. For example, the total bits of the tri-integers for the Homo sapiens globin gene is 349 bits, each nucleotide needs 2.041 bits. Therefore, encoding DNA sequences as the tri-integers may save storage space. 

\section{Conclusions}
We present a novel method for encoding a DNA sequence into three integers. Encoding a DNA sequence by integer CGR produces unique tri-integers that contain all sequence information. Therefore, the tri-integers from encoding a DNA sequence can be considered as the mathematical descriptor of the sequence. The encoding method can be a promising tool for DNA sequence compressions, encryption, and steganography.

%%%%%%%%%%%%%%%%%%%%%%%%%%%%%%%%%%%%%%%%%%%%%%%%%%%%%%%%%%%%%%%%%%%%%%%%%%%%%%%%%%%%%
%
%     please remove the " % " symbol from \centerline{\includegraphics{fig01.eps}}
%     as it may ignore the figures.
%
%%%%%%%%%%%%%%%%%%%%%%%%%%%%%%%%%%%%%%%%%%%%%%%%%%%%%%%%%%%%%%%%%%%%%%%%%%%%%%%%%%%%%%
%\section*{acknowledgement}
%There is no fund support in this project. 
%\paragraph{Funding\textcolon} 
%This research is not supported by a fund.
%\bibliographystyle{natbib}
%\bibliographystyle{achemnat}
%\bibliographystyle{plainnat}
%\bibliographystyle{abbrv}
%\bibliographystyle{bioinformatics}
%
%\bibliographystyle{plain}
%
%\bibliography{Document}

\bibliographystyle{elsarticle-harv}
\bibliography{../References/myRefs}
%\bibliographystyle{elsarticle-num}
%\bibliography{myRefs}
%% References without bibTeX database

\end{document}